\documentclass[%
 aip,
 apl,%
 amsmath,amssymb,
 reprint,floatfix%
]{revtex4-1}

\usepackage{amsmath}
\usepackage{color}
\usepackage{graphicx}
\usepackage{dcolumn}
\usepackage{bm}
\usepackage[english]{babel}

\newcommand{\CNN}{Centre de Nanosciences et de Nanotechnologies, CNRS, Universit{\'e} Paris-Sud, Universit{\'e} Paris-Saclay, Avenue de la Vauve, 91120 Palaiseau, France}
\newcommand{\IMEC}{IMEC, Kapeldreef 75, B-3001 Leuven, Belgium}

\begin{document}



\title {Size dependence of spin-torque switching in perpendicular magnetic tunnel junctions}

\author{Paul Bouquin}
 \email{paul.bouquin@u-psud.fr}
\affiliation{\CNN}
\affiliation{\IMEC}
\author{Siddharth Rao}
\affiliation{\IMEC}
\author{Gouri Sankar Kar}
\affiliation{\IMEC}
\author{Thibaut Devolder}
\affiliation{\CNN}

\begin{abstract}
We simulate the spin torque-induced reversal of the magnetization in thin disks with perpendicular anisotropy at zero temperature. Disks typically smaller than 20 nm in diameter exhibit coherent reversal. A domain wall is involved in larger disks. We derive the critical diameter of this transition. Using a proper definition of the critical voltage, a macrospin model can account perfectly for the reversal dynamics when the reversal is coherent. The same critical voltage appears to match with the micromagnetics switching voltage regardless of the switching path. 
\end{abstract}

\maketitle

Magnetization reversal in small particles is a long standing problem \cite{stoner_mechanism_1991,aharoni_nucleation_2001-1} that was recently put in a new context by the emergence of spin transfer torque (STT) magnetic random access memories \cite{khvalkovskiy_basic_2013} (MRAM). This technology is based on the STT-induced manipulation of the magnetization of nano-sized ultrathin disks with perpendicular magnetic anisotropy (PMA). In addition to its fundamental interest, the switching dynamics is of paramount application importance as it determines many of the performance metrics of this technology.  
Unfortunately experimental investigations are scarce \cite{bernstein_nonuniform_2011, gajek_spin_2012, le_gall_state_2012, sun_effect_2011,devolder_time-resolved_2016, hahn_time-resolved_2016,devolder_size_2016} probably because of the technical difficulties associated with the small dimensions as well as the large frequencies involved in the switching process. As a result the switching paths are often conjectured from reasonnable but approximate models \cite{chaves-oflynn_thermal_2015, munira_calculation_2015}, if not from overly simplified models like the macrospin picture \cite{sun_spin-current_2000, butler_switching_2012,pinna_spin-transfer_2013} whose range of validity is still to establish.

Here we unravel the size dependence of the switching dynamics by taking advantage of the accurary of micromagnetics. We first clarify how to implement STT in a way that is adequate for magnetic tunnel junctions (MTJ). We then describe the main switching regimes: coherent for small disks versus domain wall (DW) based at larger diameters. We discuss the critical diameter that separates these two regimes. We then parametrize the macrospin model to account exactly for the coherent regime. Finally we describe the size dependence of the switching voltage and provide a model valid whatever the switching mode. Our results clarify the predictive capability of the corpus of theories based on the macrospin model and therefore it has strong implications for magnetic random access memories.


\begin{table*}
\label{mytable}
\begin{tabular}{|c|c|c|c|c|c|c|c|c|}
  \hline
  Magnetization & Damping & Anisotropy & Resistance area & Tunnel & Exchange & Bloch  & Exchange  & Initial tilt \\
   &  constant & field  & product ($\theta=0$) & magnetoresistance & stiffness & length & length &   \\
  \hline
  $M_S=1.2$ MA/m & $\alpha=0.01$ &  1.566 MA/m & 8.55 $\mathrm{\Omega . \mu m^2}$ & 150\% & 20 pJ/m & 8.5 nm & 4.7 nm & $\theta_{t=0} = 1 ~\mathrm{deg.}$\\
  \hline
\end{tabular}
\caption{Material properties used in the micromagnetic simulations, meant to mimic a dual MgO FeCoB-based layer \cite{couet_impact_2017, devolder_using_2017}. 
}
\end{table*}

We are interested by the response of PMA disks to the STT associated with voltage steps applied through a tunnel junction. A first difficulty arises from the fact that the STT is most often expressed in units of current densities \cite{slonczewski_currents_2005, sun_magnetoresistance_2008} while the applied voltage is a more correct metric in a tunnel junction context. Indeed the insulating nature of the tunnel oxide renders the voltage laterally uniform across the disk, while the current density is not. To implement STT within micromagnetics, we start from the Landau-Lifshitz-Gilbert-Slonczewski equation: 
\begin{equation}
\label{eqllgs}
\dot{\boldsymbol{m}}= - \gamma \mu_0 \boldsymbol{m} \times \boldsymbol{H_{\mathrm{eff}}} + \alpha \boldsymbol{m} \times \dot{\boldsymbol{m}} + \boldsymbol{\tau _{\mathrm{STT}}}~,
\end{equation} 
where $\boldsymbol{m}$ is the normalised magnetization, $\gamma$ the gyromagnetic ratio, $\boldsymbol{H_{\mathrm{eff}}}$ the effective field, and $\alpha$ the damping constant.
We consider a spin polarization along a unit vector $\boldsymbol{p}$ parallel to the uniaxial anisotropy axis $(z)$. In this configuration, the field-like STT can be disregarded as it is mathematically equivalent to the Zeeman torque of an easy axis field. We thus reduce the STT to a sole Slonczewski-like torque $\boldsymbol{\tau_{\mathrm{Slonc}}}$. Assuming one-dimensional transport along $(z)$ we can write a local STT as~\cite{slonczewski_currents_2005}:
\begin{equation}
\boldsymbol{\tau_\mathrm{Slonc}}= \gamma\frac{\hbar}{2e\mu_0 M_s t_\mathrm{mag}} \eta(\theta) J(\theta)\boldsymbol{m} \times (\boldsymbol{m} \times \boldsymbol{p})
\end{equation}
Here $t_\mathrm{mag}=2~\textrm{nm}$ is the layer thickness, $M_s$ is its magnetization, 
and $\theta$ is the local angle between $\boldsymbol{m}$ and $\boldsymbol{p}$. The STT efficiency \cite{slonczewski_theory_2007} is $\eta = \frac{P}{1+P^2\cos(\theta)}$ where $P$ is the spin polarization \cite{julliere_tunneling_1975} linked to the tunnel magneto-resistance $\rho_\mathrm{TMR}$ ratio following  $P=\sqrt{\frac{\rho_\mathrm{TMR}}{\rho_\mathrm{TMR}+2}}$.  
Note that for simplicity we disregard the bias dependence \cite{simmons_generalized_1963,theodonis_anomalous_2006} of the MTJ conductance. With that simplifying assumption, the conductance is $G(\theta)= \frac{1+P^2\cos(\theta)}{R_\perp}$ where the median resistance 
$R_\perp=\frac{2R_{\pi} R_\mathrm{0}}{R_\mathrm{\pi}+R_\mathrm{0}}$ depends on the resistances of the $\theta=0$ and $\theta=\pi$ states. It is noticeable that the $\theta$ dependences of $J$ and $\eta$ compensate when the STT is expressed in voltage. Indeed we can write: 
\begin{equation}
\label{eqsttv}
\boldsymbol{\tau_\mathrm{Slonc}}=\gamma P \frac{V}{\mathcal{A}R_\perp}\frac{\hbar}{2e \mu_0 M_s t_\mathrm{mag}}\boldsymbol{m}\times (\boldsymbol{m} \times \boldsymbol{p}) ~,
\end{equation}
where $\mathcal{A}$ is the disk area. Equation \ref{eqsttv} recalls that the STT is symmetrical with respect to $\theta$: in the case of MTJs with neither applied field nor field-like STT, the voltage-induced $\theta=0$ to $\mathrm{\pi}$ and $\mathrm{\pi}$ to 0 transitions should occur at exactly opposite voltages and follow identical dynamics. We perform our micromagnetic simulations using MuMax3 ~\cite{vansteenkiste_design_2014}. In this software the implementation of Eq.~\ref{eqsttv} requires to set $\epsilon '=0$, $P_\mathrm{mumax}=P$, $\Lambda=1$ and $J_\mathrm{mumax}=\frac{V}{\mathcal{A}R_{\perp}}$(see sec. III.H in ref.~\onlinecite{vansteenkiste_design_2014}). For numerical accuracy, the cell size is kept below $2 \times 2~\mathrm{nm^2}$, i.e. substantially smaller than the characteristic micromagnetic lengths of our magnetic material (Table I). 


\begin{figure*}[t!]
\begin{center}
\includegraphics[width=17cm]{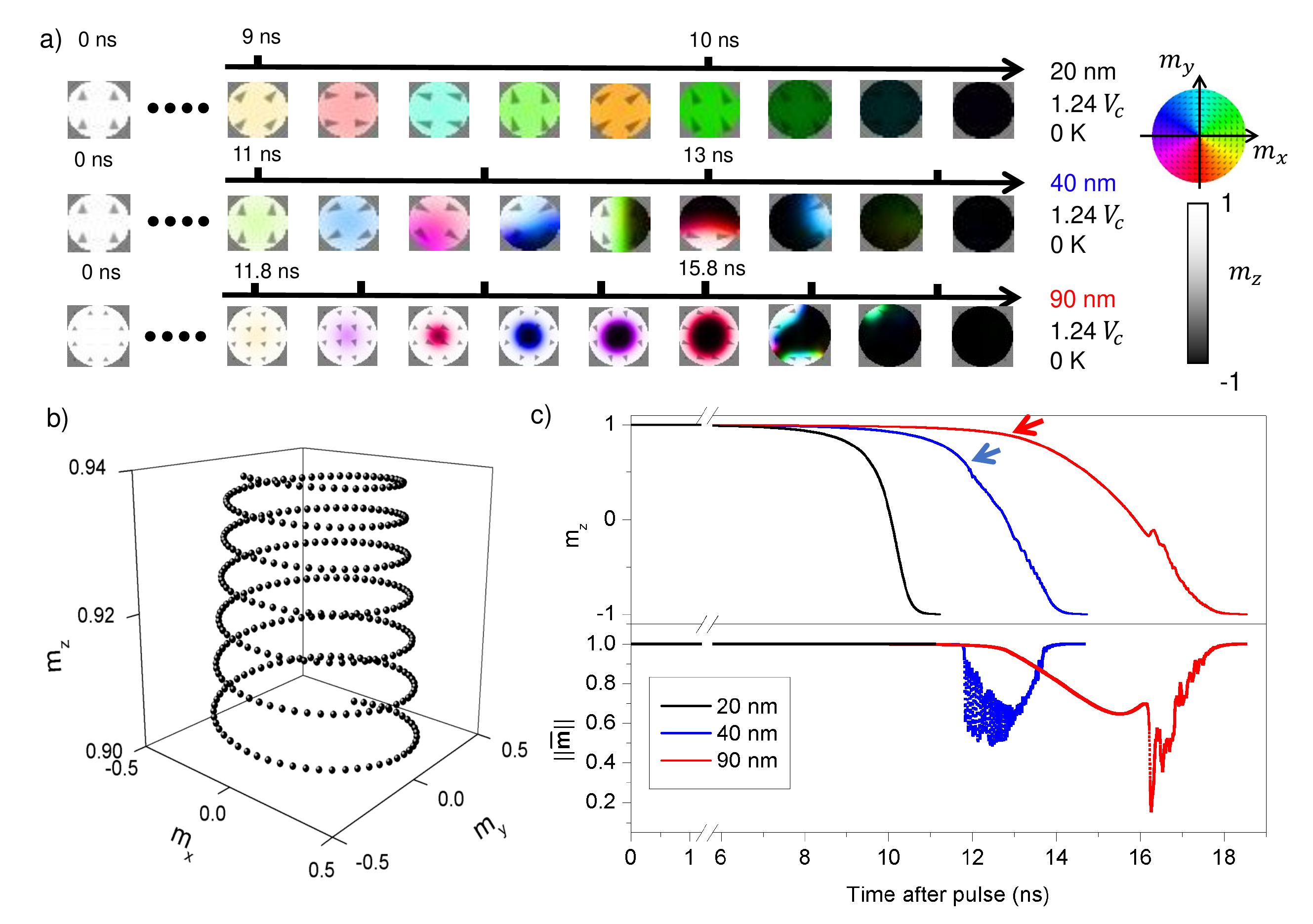}
\caption{Size dependence of the switching path at zero temperature. a) Snapshots of the magnetization during the switching for different sizes. b) Typical magnetization trajectory when the reversal is coherent. c) Mean value of the $z$ component of the magnetization and modulus of the mean magnetization during the reversal for 3 different sizes.}
\end{center}
\end{figure*}

Let us study the size dependence of the switching path at zero temperature. In order to better evidence the influence of the diameter, we apply voltages that correct for the slight dependence of the switching voltage over the junction diameter; the exact procedure will be detailed later. We varied the applied voltage between $5 \ \%$ and $50 \ \%$ above the normalized critical switching voltage and found that increasing the voltage accelerates the dynamics without altering the nature of the switching path (not shown). We varied the diameter between 16 nm and 300 nm. The reversal is coherent below 22 nm while a (DW) is involved for  disks larger than 26 nm until complexity substantially grows for diameters above 50 nm with the appearance of a center domain. The diameters of 20 nm, 40 nm and 90 nm (Fig. 1) illustrate those three regimes. 

For the 20 nm disk, the magnetization remains uniform all along the reversal (Fig. 1.a). The degree of coherence during the reversal can be measured by the modulus of the mean magnetization $||\boldsymbol{\bar{m}}||=\sqrt{m_x^2+m_y^2+m_z^2}$ where $m_i$ is the spatial average of the $i$ component of $\boldsymbol{m}$. For 20 nm and below, there is no perceivable loss in the degree of coherence (Fig. 1.c, black curve): the magnetization switches through a gradual decrease of its $z$ component while the in-plane components precess in quadrature (Fig. 1.b). For a 22 nm disk the reversal starts to exhibit a faint transient non-uniformity while the reversal remains mostly coherent and precessional (not shown). A further increase of the diameter leads to a gradually stronger non-uniformity until a 180 degree DW appears during the reversal for disks larger than 26 nm. This DW based reversal is exemplified with the 40 nm disk (Fig. 1.a): the reversal starts by a coherent phase lasting 11 ns during which the magnetization undergoes a growing precession that recalls the behavior seen for smaller disks. During this coherent initial phase, the $m_z$ component of the magnetization is slightly non-uniform in the sense that the precession cone is more opened at the center than near the disk circumference. The larger susceptibility near the disk center can be understood from the demagnetizing field profile, which is maximal at the disk center and therefore reduces locally the effective anisotropy. 
In the 26 to 60 nm disks, once a large precession amplitude is reached, a nucleation occurs which leads to the creation of a wall near (but not from) an edge of the disk as noticed already in ref.~\onlinecite{munira_calculation_2015, you_switching_2014}. Note that the DW is a genuine 180 deg. wall: it separates regions with $\theta=0$ and $\theta=\pi$. We emphasize that despite the magnetization being tilted everywhere before the nucleation, the nucleation projects the magnetization to either of the two easy directions, with then no remaining tilt in the domains after the DW creation. The DW then sweeps across the disk in a non-trivial manner until saturation. This DW creation slows down the decay of $m_z$ (see the sudden decrease of the slew rate of $m_z(t)$, blue arrow in Fig. 1.c). 
Above diameters of 60 nm, the radial gradient of the precession cone during the initial coherent phase of the reversal gets even more pronounced such that the magnetization at the center dips and a reversed domain is formed at the center. The formation of the central domain is a gradual process (no change in topology) that does not lead to any specific feature in the $m_z(t)$ curve (red arrow in Fig. 1.c). 
Once created, the central domain expands till the disk edges; this lasts longer for larger devices. Once the central domain approaches the disk edge, it wets one or several points of the perimeter of the disk, depending on the disk size. This wetting is sudden and has a clear signature in the $m_z(t)$ curve.


\begin{figure}[t!]
\begin{center}
\includegraphics[width=8.5 cm]{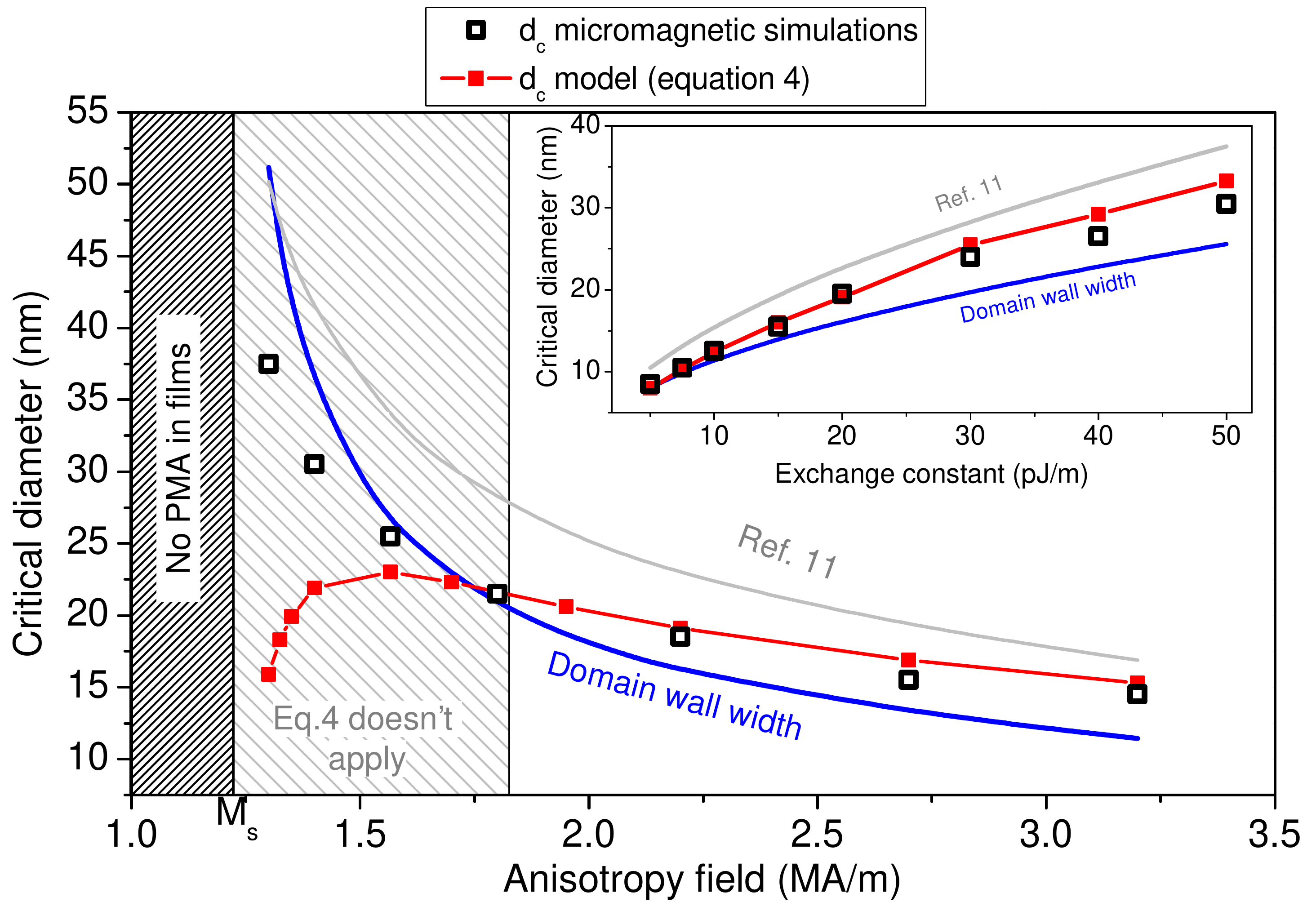}
\caption{Critical diameter above which a 180 deg. DW is observed in the micromagnetic simulations and comparison to its analytical estimates. Main panel: while varying anisotropy field at fixed $A_{ex}=20 \ \mathrm{pJ} / \mathrm{m}$ or while varying exchange stiffness at fixed $H_{k}=2.2 \ \mathrm{MA} / \mathrm{m}$ (inset).}
\end{center}
\end{figure}

Let us focus on a central result from this study: the critical diameter $d_c$ above which the reversal involves DWs. Several expressions were proposed in the literature to predict $d_c$. In the minimal approach\cite{chaves-oflynn_thermal_2015, khvalkovskiy_basic_2013}, $d_c$ is estimated by comparing two energies. First the energy cost of placing a DW along the diameter of the disk. In ref.\cite{chaves-oflynn_thermal_2015, khvalkovskiy_basic_2013}, this energy is written $4 \sqrt{A_\mathrm{ex} K_\mathrm{eff}^\mathrm{disk}}d \ t_\mathrm{mag}$ where $A_\mathrm{ex}$ is the exchange stiffness and $K_\mathrm{eff}^\mathrm{disk}$ the effective anisotropy of the disk. These expressions assume that the disk is larger than the domain wall i.e. $d>\pi \Delta $ where $\Delta=\sqrt[]{\frac{A_\mathrm{ex}}{K_\mathrm{eff}^\mathrm{film}}}$. The second relevant energy is the one of the system when the magnetization is uniformly in-plane:  $\frac{\pi}{4}K_\mathrm{eff}^\mathrm{disk}d^2t_\mathrm{mag}$ which is necessary to overcome for a coherent reversal. The effective anisotropy of the disk depends on the demagnetizing factors $N_z(d)$ and $N_x(d)$. 
 We deduce $N_z-N_x$ with micromagnetic simulations from the FMR angular frequency using $ H_\mathrm{k,eff}^\mathrm{disk}= H_k - (N_z-N_x) M_s = \frac{\omega_\textrm{FMR}}{\gamma \mu_0} $. 
 Solving self-consistently this minimal approach would yield a critical diameter of 33.6 nm which is larger than the DW width but substantially differs from the 26 nm observed in micromagnetics.

This difference can result from one fundamental and two technical deficiencies of the minimal approach. Fundamentally, any comparison based on the energies of static configurations is bound to underestimate the energy cost of an inherently dynamical process like reversal. However in STT switching, the energy lost by damping is supposed to be compensated by STT; therefore we conjecture that we can overlook this fundamental objection.  Besides, the minimal approach confuses the DW energy within a disk with that within a fictitious infinite film that would have the same effective anisotropy as the disk. Finally it neglects the  dipolar energy gained when breaking the system into domains. 
Neglecting the domain-to-domain dipolar couping is a minute error in the ultrathin limit \cite{hubert_domain_1998}. Conversely the imprecision in DW energy can be substantial. Indeed the demagnetizing field in thin films is essentially local within a DW \cite{thiaville_dynamics_2012}, such that the wall energy is much more linked to the effective anisotropy of the film rather than the effective anisotropy of the disk, and therefore should be taken as $4 \sqrt{A_\mathrm{ex} K_\mathrm{eff}^\mathrm{film}}d \ t_\mathrm{mag}$.

Alltogether, an improved estimate of $d_c$ is
\begin{equation}
\label{eqdc}
d_c\approx \frac{16}{\pi} \frac{\sqrt{A_\mathrm{ex} K_\mathrm{eff}^\mathrm{film}}}{K_\mathrm{eff}^\mathrm{disk}}=\frac{16}{\pi} \sqrt{\frac{2 A_\mathrm{ex}}{\mu_0 M_s}} \times \frac{ \sqrt{H_k- M_s}}{H_k-(N_z - N_x)M_s}
\end{equation}
which also has to be solved self-consistently because $(N_z - N_x)$ depends on $d$.  Fig. 2 compares Eq.~\ref{eqdc} with the micromagnetic $d_c$ in a relevant interval of anisotropy field and exchange stiffness. The matching is satisfactory: as long as the domain wall width exceeds the disk diameter, Eq.~\ref{eqdc} is reliable estimation of $d_c$. Note that for a fast evaluation of eq.~\ref{eqdc}, the demagnetizing factors can be estimated from Eq. 11 of ref.~\onlinecite{beleggia_equivalent_2006} for $t_\mathrm{mag}/d > 0.06$ and from  ref.~\onlinecite{mizunuma_size_2013} otherwise.



The switching dynamics in the macrospin model has been used extensively in the past to predict the switching speeds \cite{sun_spin-current_2000}, the error rates \cite{butler_switching_2012,tzoufras_switching_2017} and the stability diagrams \cite{le_gall_state_2012}. In order to assess to predictive ability of these studies, it is important to evaluate to what extent the macrospin model can be used to mimic the coherent reversal regime. We describe the macropin with its cylindrical coordinates $m_z$ and $\phi$. The LLGS equation of a macrospin reduces \cite{pinna_spin-transfer_2013, butler_switching_2012} to:
\begin{eqnarray}
\dot{m_z} = \frac{1}{\tau}(m_z + v)(1-m_z ^2) \label{eqmz}\\
\dot{\phi} = \frac{1}{\alpha \tau}(m_z + \alpha^2 v) \label{eqphi}
\end{eqnarray}
where $\tau=\frac{1+ \alpha ^2}{\alpha \gamma \mu_0 H_\mathrm{k,eff}^\mathrm{disk}}$ 
and $v=\frac{V}{V_{c}}$ is the voltage normalized to the macrospin critical voltage\cite{butler_switching_2012}, i.e. the smallest voltage at which the $\theta=\pi$ state gets unstable: 
\begin{equation}
\label{eqvc}
V_{c}= \frac{2 \alpha e \mathcal{A}R_\perp t_\mathrm{mag} \mu_0 M_s H_\mathrm{k,eff}^\mathrm{disk}}{P \hbar} 
\end{equation}
Two words of caution are needed. First, although here we restrict to the case of zero applied field, in general the time evolution of $m_z$ is triggered by all torques. As a result some authors \cite{butler_switching_2012, tomita_unified_2013} choose to aggregate the external field $H_z$ and the spin-torque by defining a \textit{generalized} stimulus and substituting $\frac{V}{V_{c}}$ by $\frac{V}{V_{c}} + \frac{H_z}{H_{c}}$ in Eq.~\ref{eqmz}. We prefer not to perform this substitution in Eq.~\ref{eqmz} because the same substitution cannot be applied to Eq.~\ref{eqphi}: qualitatively, the influence of Zeeman torque onto the precession frequency (Eq.~\ref{eqphi}) is way higher that of the STT. $H_z$ and $V$ can't be aggregated when describing the precession. 
Secondly, we stress that since the macrospin is meant to mimic a disk, its $H_\mathrm{k,eff}^\mathrm{disk}$ must take into account the demagnetizing term $-(N_z -N_x)M_s$. 


Fig. 3 compares the switching dynamics obtained for the macrospin and for the largest disk showing coherent switching in micromagnetics. Provided the proper $H_\mathrm{k,eff}^\mathrm{dev}$ and the resulting adequate $V_c$ are used, the outcomes of the two models match for the time evolution of $m_z$ (Eq.~\ref{eqmz}, Fig. 3.a) as well as for the instantaneous precession frequency (Eq.~\ref{eqphi}, Fig. 3.b). For larger disks the perfect match is maintained during the initially coherent phase of the reversal (not shown), but as expected the macrospin model fails to account for the subsequent evolution as soon as a non-uniformity sets in (not shown).



\begin{figure}[t!]
\begin{center}
\includegraphics[width=8.5 cm]{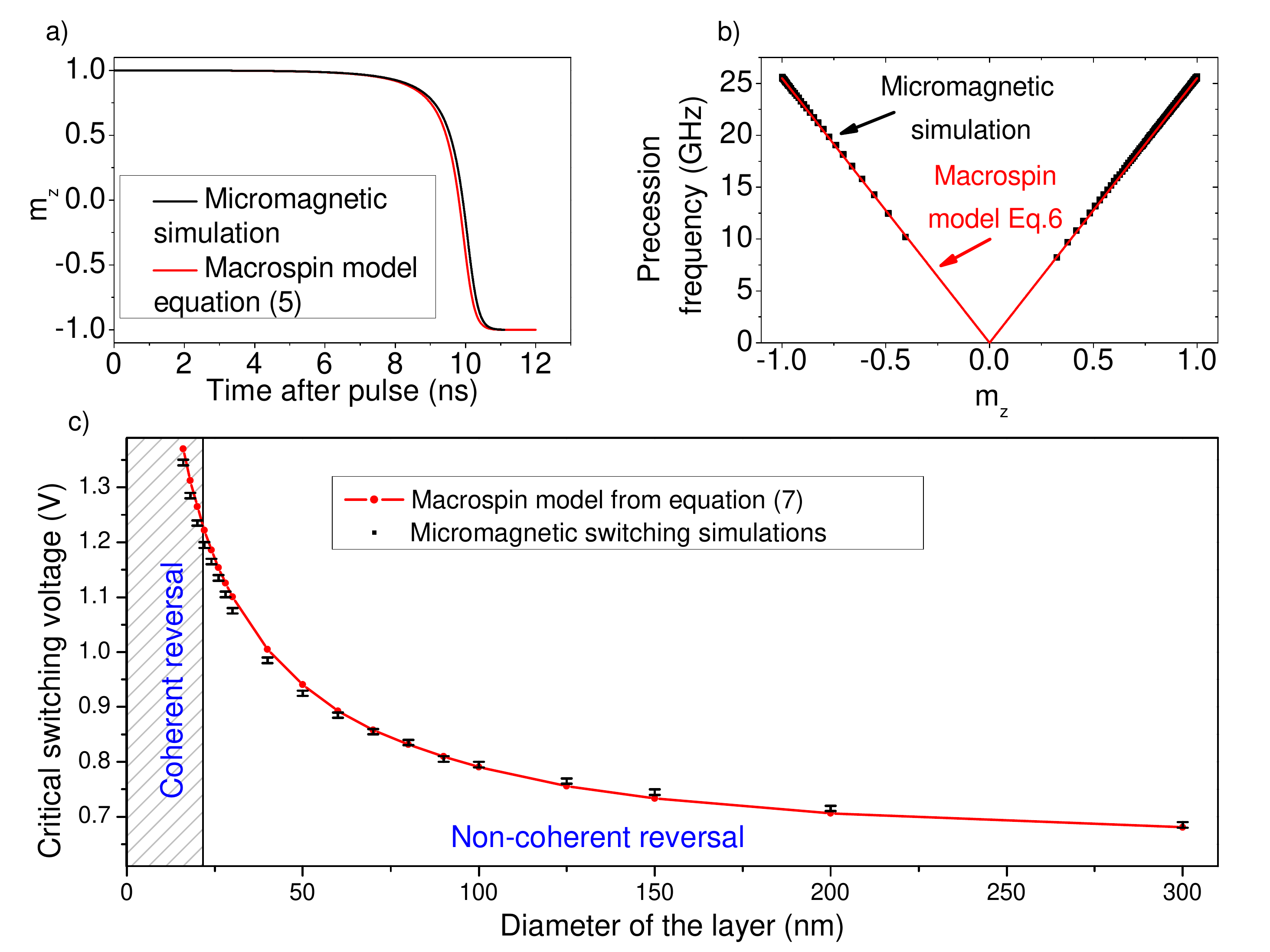}
\caption{Comparison between micromagnetic simulations and macrospin model for a 20 nm disk at 0 K under 1.24 $V_c$: a) Mean value of the out plane component of the magnetization $m_z$. b) $m_z$ dependence of the instantaneous precession frequency. c) Comparison of the macrospin critical voltage and the micromagnetic switching voltage for different diameters.}
\end{center}
\end{figure}

From the previous discussion, we conclude that the macrospin model describes perfectly the coherent reversal regime. Let us now see whether the macrospin critical voltage (Eq. \ref{eqvc}) can account for the micromagnetic switching voltage, including for sizes that lead to non-coherent reversal. 
The voltage that leads to a destabilization of the uniformly magnetized state is bound to match the macrospin critical voltage $V_c$. However destabilizing the uniformly magnetized state is a \textit{necessary} condition for switching but it might not be a \textit{sufficient} condition. Indeed even if uniform state is unstable to finite amplitude precession, the precession amplitude can be limited by non-linearities and not lead to reversal, as observed in in-plane magnetized metallic spin-valve \cite{Kiselev_microwave_2003, devolder_instability_2005}, in which there is a net difference between instability and switching. In our case we find that $V_c$ and the micromagnetic switching voltage do agree for all investigated disk diameters (Fig. 3.c). 
This indicates that destabilizing the uniformly magnetized state is a necessary \textit{and} sufficient condition for switching, and that this holds even when the reversal is far from coherent. In short, Eq. \ref{eqvc} is the switching voltage including for sizes that lead to non-coherent reversal.

In summary, we have simulated the spin-torque induced switching of the magnetization of disks with perpendicular anisotropy. The reversal always starts by the amplification of a circular precession. For disk diameters below a critical threshold, the reversal is coherent and can be accounted for by a macrospin model. For larger sizes a domain wall appears during the reversal. Energy considerations can predict this critical size. Besides, the macrospin critical voltage (Eq.~\ref{eqvc}) predicts the switching voltage for any sizes, including when the reversal is not coherent.
This work was supported by IMEC's Industrial Affliation Program on the STT-MRAM devices.

\bibliography{biblio_zotero}

\end{document}